\documentclass[letterpaper]{article}
\usepackage{flairs}
\usepackage{times}
\usepackage{helvet}
\usepackage{courier}
\usepackage{multirow}
\usepackage{blindtext}
\usepackage[ruled,vlined, linesnumbered]{algorithm2e}
\usepackage{adjustbox}
\usepackage{amsfonts}
\usepackage{xspace}

\begin{document}

\title{IoTFlowGenerator: Crafting Synthetic IoT Device Traffic Flows for Cyber Deception}
\author{Joseph Bao\\ The University of Texas \\ at Dallas \\ jxb110430@utdallas.edu \\ \And  Murat Kantarcioglu\\ The University of Texas \\ at Dallas \\ muratk@utdallas.edu \\ \And Yevgeniy Vorobeychik \\ Washington University \\ in St. Louis \\ yvorobeychik@wustl.edu
\And Charles Kamhoua\\
Army Research Labs \\ charles.a.kamhoua.civ@\\army.mil}

\maketitle
\date{}
\begin{abstract}
Over the years, honeypots emerged as an important security tool to understand attacker intent and deceive attackers to spend time and resources. Recently, honeypots are being deployed for Internet of things (IoT) devices to lure attackers, and learn their behavior. However, most of the existing IoT honeypots, even the high interaction ones, are easily detected by an attacker who can observe honeypot traffic due to lack of real network traffic originating from the honeypot. This implies that, to build better honeypots and enhance cyber deception capabilities, IoT honeypots need to generate realistic network traffic flows.

To achieve this goal, we propose a novel deep learning based approach for generating traffic flows that mimic real network traffic due to user and IoT device interactions.
A key technical challenge that our approach overcomes is scarcity of device-specific IoT traffic data to effectively train a generator.
We address this challenge by leveraging a core generative adversarial learning algorithm for sequences along with domain specific knowledge common to IoT devices.
Through an extensive experimental evaluation with 18 IoT devices, we demonstrate that the proposed synthetic IoT traffic generation tool significantly outperforms state of the art sequence and packet generators in remaining indistinguishable from real traffic even to an adaptive attacker.
\end{abstract}

\section{Introduction}

Deception has been prevalent throughout history, especially in warfare. 
In the context of cybersecurity, various cyber deception techniques have been applied to move the attackers away from the real resources and learn their intent. An important cyber deception technique that has been heavily used is honeypots. Honeypots have been deployed to deceive attackers into spending time and resources away from real targets and disclose their tools and capabilities.
After many attacks against IoT devices, honeypot based solutions have been applied in the  context of Internet of Things (IoT) devices as well.

Although existing IoT honeypots have significant capabilities, what remains a major weakness is their inability to generate network traffic that appears sufficiently realistic in such a way that it cannot be easily attributed to a honeypot by attackers.
This is especially important for IoT deployment settings where an attacker can observe some of the network traffic. 
For example, in a commercial IoT or Industrial Internet of Things (IIoT) deployment setting, once an attacker hacks into a local network and starts observing the network, it can easily detect IoT/IIoT honeypots since they will not have a traffic flow originating from them. Therefore, to create IoT honeypots that are difficult to detect, there is a need to generate realistic network traffic flows associated with those honeypots.

A simple idea for generating such synthetic traffic flows could be to replay captured existing IoT device traffic.  However, replay is easy to detect as long as the adversary is observing traffic for a sufficiently long period of time.
Another option could be to use state of art network traffic generators that leverage generative adversarial network (GAN) frameworks. However, direct application of these techniques in our setting does not result in effective generators. 


In this work, we propose a \textit{generative IoT network traffic flow data algorithm that can generate network traffic using a small amount of real IoT data}. In addition, the developed generator can deceive the attacker into accepting honeydata with much higher probability compared to the state of art network generator approach. 
Our contributions in this work can be summarized as follows: 

\begin{itemize}
\item We develop a novel IoT traffic flow generation algorithm that can generate realistic synthetic traffic flow \footnote{We also use the term honeydata for such synthetic IoT traffic flows.} for cyber deception.
\item Using signature extraction, our algorithm can easily learn how to generate realistic network traffic with access to little real IoT data.
\item By comparing our algorithm with a state of art network traffic generator using 18 different IoT devices, we show that our algorithm performs much better in an adversarial setting.  
\end{itemize}

In the remainder of this paper, we first discuss related work in section~\ref{sec:background}, we provide the necessary background related to our algorithm. In section~\ref{threatmodel}, we discuss our threat model assumptions. In section~\ref{sec:details}, we present the implementation of our synthetic IoT traffic generation algorithm. In section~\ref{sec:results}, we show the experimental evaluation results using different IoT devices.
We conclude the paper in section~\ref{sec:conclusion}.

\section{Background}
\label{sec:background}
In this section, we introduce two state of art Generative Adverserial Networks: SeqGAN, which utilizes reinforcement learning concepts to generate sequences in a discrete space, and DoppelGANger, a state of the art framework for generating time series datasets with metadata - including network traffic. We use SeqGAN as part of our IoT network traffic flow generation algorithm, and use DoppelGANger as a comparison model to evaluate the relative fidelity of our generated traffic flows.

\subsection{Generative Adverserial Networks}
Generative Adversarial Networks that use discriminative models to guide the training of the generative model have enjoyed considerable success in generating real-valued data. GANs have many applications such as face synthesis \cite{DBLP:journals/corr/abs-1910-14247}, face aging \cite{antipov2017face,zhang2017age,8578926}, image blending \cite{wu2019gpgan,Chen2019TowardRI}, text synthesis \cite{xu2017attngan,zhang2017stackgan}, object detection \cite{ehsani2018segan,li2017perceptual,Bai2018SODMTGANSO}, language and speech synthesis \cite{lin2018adversarial,hsu2017voice}, music generation \cite{mogren2016crnngan,guimaraes2018objectivereinforced}, etc. Even though GANs have achieved some great results by producing highly realistic samples, it is still difficult to train GANs with good stability. Other important shortcomings include mode collapse \cite{goodfellow2017nips,arora2017generalization}, vanishing gradient \cite{goodfellow2017nips}, and the lack of proper evaluation metrics \cite{borji2018pros}. For IoT traffic, or sequential data, GAN's have limitations when the goal is to generate sequences of discrete valued data.  SeqGAN is a GAN based sequence generation framework that addresses this problem~\cite{DBLP:journals/corr/YuZWY16}. Modeling the data generator as a stochastic policy in reinforcement learning (RL), SeqGAN bypasses the generator differentiation problem by directly performing gradient policy update. The reinforcement learning reward signal comes from the GAN discriminator judged on a complete sequence, and is passed back to the intermediate state-action steps using Monte Carlo search. Extensive experiments on synthetic data and real-world tasks demonstrate significant improvements over strong baselines. \textit{SeqGAN alone is insufficient for generating synthetic network traffic and is limited to generating only uni-variate integer sequences}. Our work extends the work of SeqGAN by generating multi-variate sequences composed of discrete and continuous values.


DoppelGANger is a synthetic generation framework for time series datasets including network traffic~\cite{DBLP:journals/corr/abs-1909-13403}. 
To model correlations between measurements and their metadata, DoppelGANger decouples the generation of metadata from time series and feeds metadata to the time series generator at each time step. The DoppelGANger framework allows generation of multi-variate time series datasets with minimal expert knowledge. We use DoppelGANger to generate synthetic IoT network traffic to serve as a baseline predictive model to draw comparisons from.


\subsection{VQ-VAE}
Variational AutoEncoders (VAEs) are well-studied generative models. VAEs have been applied to various generation tasks. It is generally easier to train VAEs than GANs. However, GAN-based methods have been able to generate high-fidelity synthetic images, whereas, until recently, VAE-based methods
seemed to generate blurred images. In order to address this issue of low-quality generation, Van Den Oord et al. \cite{DBLP:journals/corr/abs-1711-00937} recently proposed the Vector Quantized-Variational AutoEncoder (VQ-VAE) framework. Unlike a regular VAE which encodes the input in a continuous latent space, VQ-VAE finds compressed representations of images by projecting the output of the encoder onto a discrete latent space. The output of the encoder is quantized to a set of vectors, called the codebook which is a priori fixed and is much smaller than the dimension of the input data. The vectors (codes) in the codebook are updated throughout the training process and VQ-VAE uses this codebook to obtain a discrete, or “vectorized”, latent representation of the output of the encoder by searching the nearest element in the codebook. Although VQ-VAE was originally designed to handle both image and audio data, it was recently extended to other natural language processing tasks. \textit{VQ-VAE is utilized in our work to extend SeqGAN to generate multi-variate sequences composed of discrete and continuous values.}

\section{Threat Model} \label{threatmodel}
The goal of our adversary is to identify active IoT device types from captured traffic in a target smart environment when the IoT device traffic is encrypted. 
We assume in our threat model, the attacker passively sniffs traffic between the Internet Gateway and IoT devices, and uses analysis techniques trained with previously obtained datasets to determine device types that exist. The term device type in this work is defined to represent the combination of make, model and software version of a particular device. Our goal of generating synthetic traffic is to trick our adversary into believing the fake traffic flow data is real and belongs to a particular device type. \par

\subsection{Adversary Assumptions}
We make the following assumptions about the adversary for our evaluation task. 

\begin{itemize}
    \item The adversary passively sniffs the network traffic and does not interact with any devices.
    \item The adversary is assumed to only have \textit{access to encrypted traffic flows} corresponding to individual device types. In other words, the content of the network packets are protected by the secure encryption.
    \item The adversary has access to protocol headers data on all layers that are not protected by the encryption scheme.
    \item The adversary makes a classification decision after witnessing $L$ packets of any given flow $f$ where $L$ can be adapted based on the attack setting.
\end{itemize}

\subsection{Adversarial ML Model}
We create an adversarial attack model using packet metadata inference, which is an adversarial ML model that classifies device traffic using raw packet metadata. We utilize packet metadata inference to evaluate the quality of generated synthetic data.

\subsubsection{Packet Metadata Inference}
The packet metadata adversary uses packet frame lengths, packet timing, packet direction, protocol configuration, source and destination port of each packet in the traffic window for inference. Each unique packet frame length, port number, direction and protocol configuration are categorically encoded, while timings are continuously encoded. These features are used as input to an adversarial classification model.
This adversarial model is a deep learning model that uses an LSTM hidden layer, which is well suited to handle sequential data. The LSTM architecture has an internal mechanism called gates that can regulate the flow of information, and can learn which data in a sequence is important to keep or throw away. By doing that, it can pass relevant information down the chain of sequences to make predictions. The model also uses a cross entropy loss function. Our results show that that this adversarial ML model is also effective for identify device types from captured encrypted traffic.

\section{IoTFlowGenerator}
\label{sec:details}

\subsection{Overview}
\begin{figure*}[htp]
    \centering
    \includegraphics[width=\textwidth]{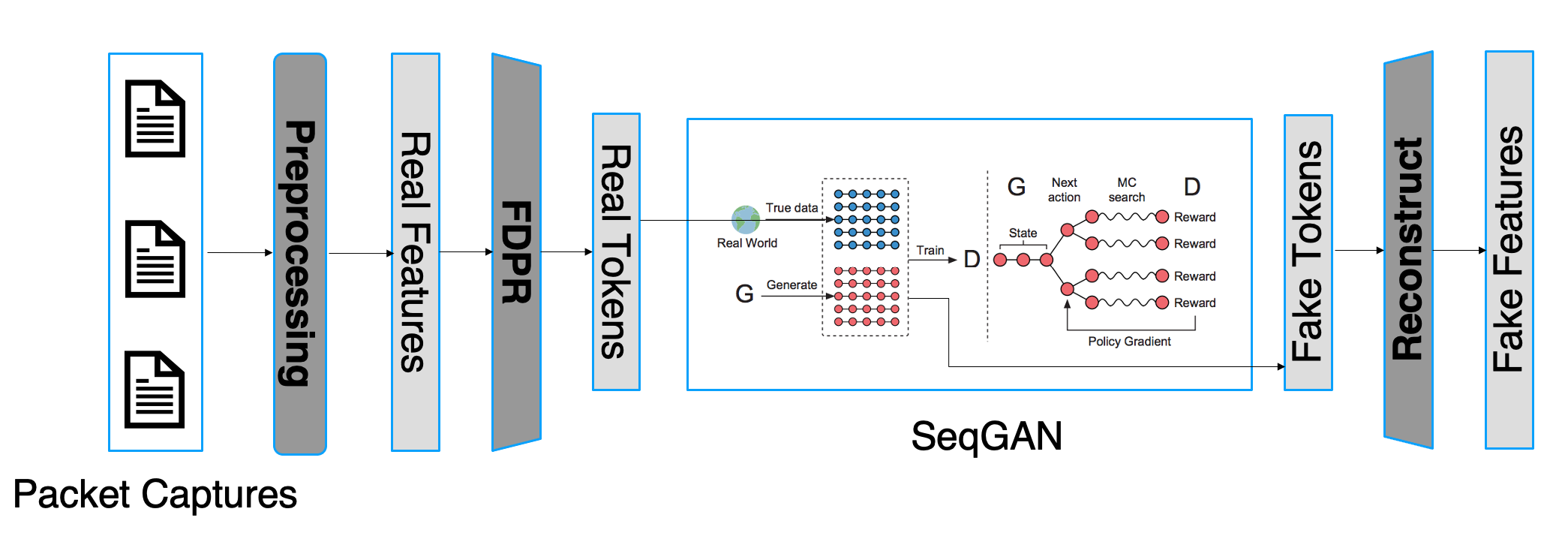}
    \caption{The Illustration of IoTFlowGenerator: Metadata is extracted from packet capture files and preprocessed into feature vectors. IoTFlowGenerator then finds discrete packet representations (FDPR) for each packet. SeqGAN is utilized to adverserially train on the real tokens and synthesize fake tokens. Finally, IoTFlowGenerator reconstructs packet feature vectors for the fake tokens, producing our desired synthetic feature vectors.}
    \label{fig:IoTFlowGenerator}
\end{figure*}

IoTFlowGenerator, as shown in Figure~\ref{fig:IoTFlowGenerator} reduces the dimensionality of traffic flows by finding discrete token representation for each packet feature vector and learning a reconstruction process to create a two way mapping between the feature space and the discrete space. The discrete packet representations enable the adversarial network to learn the temporal features of the sequential data in a lower dimensional space. At the core of IoTFlowGenerator, is SeqGAN, the process responsible for generating sequences of discrete tokens at this lower dimensional space. IoTFlowGenertor finishes by reconstructing the synthetic token sequences back into the multivariate sequential input. Below we discuss the different components of IoTFlowGenerator.

\subsection{Data Pre-processing} \label{dataprep}
As the traffic flows are initially in the form of PCAP files from our datasets, we must transform them into something that can be more easily analyzed using statistical methods.

\subsubsection{Packet Analysis via Wireshark}
We use Wireshark to analyze and extract relevant information from our datasets of PCAP files. Wireshark is a popular network protocol analyzer, and has tools for capturing, inspecting, and analyzing data packets. We deployed a script that uses Wireshark's command line utility (tshark) to parse packet capture files from our datasets and extract useful features.

\subsubsection{Data Extraction}
For each packet, we use wireshark to extract the packet source and destination, the timestamp, packet length , source and destination port number, and protocol configuration (\textit{i.e., the information available to adversaries even if the payload is encrypted}). We then convert the timestamps into a duration value measuring time in seconds until the next packet in the same capture, after all packets in the capture are sorted by earliest timestamp. We assign a direction to each packet based on an inferred target device IP and discard the IP addresses.

\subsubsection{Protocol Configurations}
We use a binary sequence to represent the presence of protocols in packet headers. We select some protocol types Wireshark can detect, namely, ARP, LLC, IP, ICMMP, ICMPv6, EAPOL, TCP, UDP, HTTP, HTTPS, DHCP, BOOTP, SSDSP, DNS, MDNS, and NTP.

\subsubsection{Assigning Packet Direction}
Most IoT devices send packets to a service provider (e.g., for using a voice-command detection module running the cloud) and receives responses from the cloud. Hence, separating packets going in different directions can be useful (e.g., to the service provider vs. from the service provider).  In order to assign a direction to a packet, an IP address of the target device is needed. For our purposes, we used the most common IP address as the IP of the target device. Each packet is given a direction, outgoing if the source IP of the packet and the IP of the target device are equal, incoming if otherwise. This method works even if there is no available data on the target device's IP address. For the purposes of this paper, the direction derived from this method was useful for our downstream prediction tasks. The IP addresses are discarded after the packet directions are obtained.

\subsubsection{Traffic Windows}
\begin{figure}[htp]
    \centering
    \includegraphics[width=8cm]{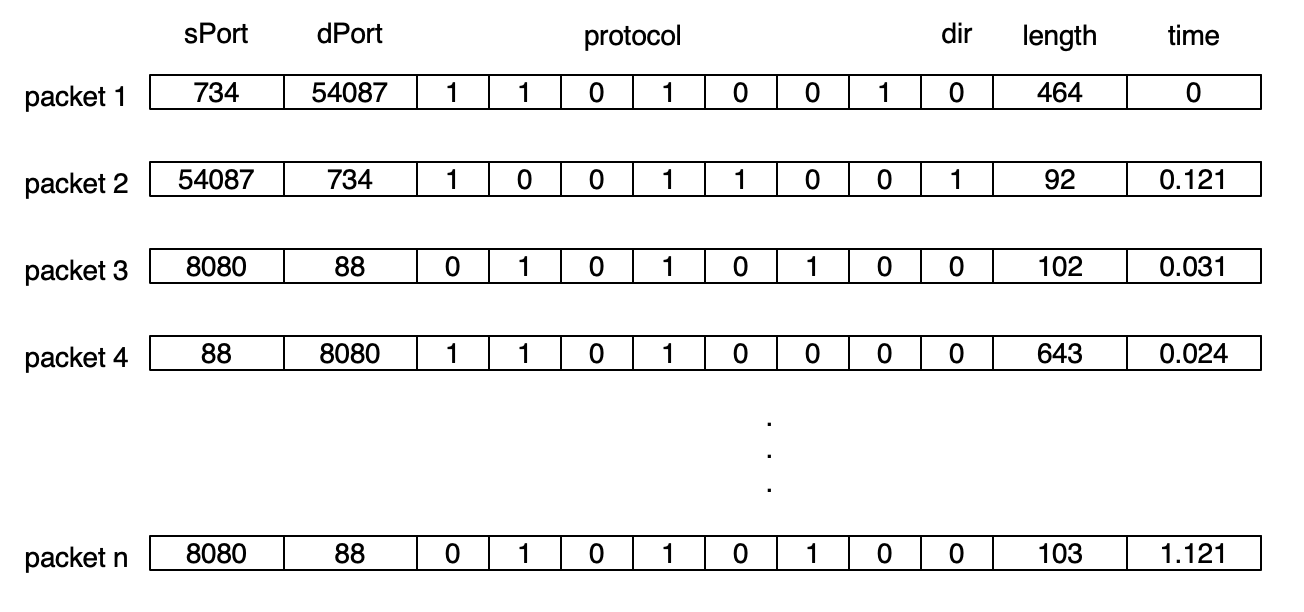}
    \caption{A traffic window with size \textit{n}}
    \label{fig:trafficwindow}
\end{figure}

Our datasets contained packet metadata sequences of variable length, but our sequential generation algorithms is currently limited to generating fixed length sequences. Thus, we group $L$ consecutive packets to form a traffic window as shown in figure \ref{fig:trafficwindow}. These traffic windows, simply fixed length device traffic flows, are useful in downstream inference tasks. For each device, we randomly select $n$ traffic windows from the device traffic flows which are used for all downstream tasks, including generation and evaluation tasks; these tasks expect fixed length input.

\subsubsection{Packet-Level Signature Extraction}
\label{signatureextraction}
One insight discovered for IoT devices, is that clearly visible changes in traffic rates directly correspond with user activities \cite{DBLP:journals/corr/abs-1812-00955}, and that traffic flows that occur immediately after certain user activities were typically comprised of packets going in opposite directions \cite{DBLP:journals/corr/abs-1907-11797}. These traffic flows for specific user activities are called packet-level signatures that consist only of the lengths and directions of a few packets in the captured traffic and can be used to infer the device type and the specific type of activity that occurred. Packet level signatures are extracted using unsupervised learning methods as shown in appendix figure \label{patternextraction}. The Density-Based (DBScan) clustering algorithm is used with an Euclidean pairwise distance metric to find signatures of arbitrary size. For each possible signature size $i$, all subarrays of length $i$ in the device type's traffic flows are used as input to the DBScan clustering algorithm. The subarrays contain only the packet length and direction, and not the duration or timing information. For clustering, we configure the maximum distance $d$ between two samples for one to be considered in the neighborhood of the other, and the minimum samples $s$ required to form a cluster. If two samples have packets at the same index with opposing directions, the distance measure between these two samples is considered maximal. For each cluster discovered and each subarray index, a range of integer packet lengths with direction encodes the set of potential packet lengths and the direction at that position. This range is determined by the minimum and maximum packet lengths in each cluster and index. A sequence of these ranges is a \textbf{packet-level signature}, and can match at any index position in a traffic flow.

\begin{figure*}[htp]
    \centering
    \includegraphics[width=\textwidth]{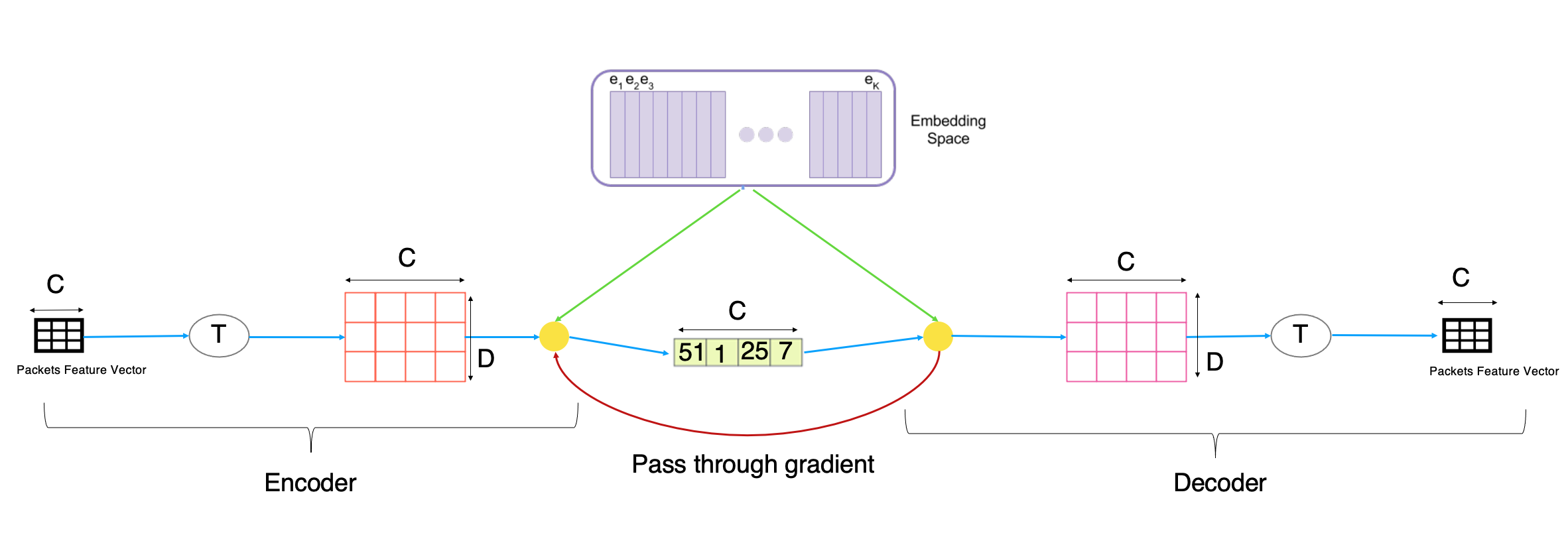}
    \caption{The Illustration of VQ-STAE: The auto-encoder used in IoTFlowGenerator to map between the feature space and the discrete space. VQ-STAE is a modified Vector Quantized - Variational Auto Encoder that uses multi-variate sequence transformers for the encoder and decoder to train on multi-variate sequences in a self-supervised manner. In the figure, $C$ represents the input sequence length and $D$ represents the length of the embedding vectors.}
    \label{fig:VQSTAE}
\end{figure*}

\subsubsection{Packet Frame Lengths and Direction}
If each packet frame length and direction was directly encoded, we encounter the curse of dimensionality when doing further analysis. To avoid this, extracted packet-level signatures are utilized to reduce the dimensionality of packet frame length and direction features. 
Our proposed model, IoTFlowGenerator uses extracted packet-level signatures to convert a traffic window into a sequence of non-overlapping activities (from extracted signatures) and orphans, namely packets that do not belong to an extracted signature. Since multiple signatures may overlap, we sort packet-level signatures by total number detected over the entire set of traffic windows for the specified device type multiplied by the length of the packet-level signature. This is because packet-level signatures that appear more frequently, are more important to consider. Each traffic window is then fully traversed using a recursive algorithm range matching the signatures in sorted order. Each sequence element can be assigned a single signature, and some elements may not be assigned a signature at all - orphan packets. Each signature is split into individual packet length ranges, one for each index position, and each unique packet in the traffic window is assigned an unique integer encoding we call packet frame tokens based on the signature and index it belongs to or the packet frame length if it does not have a signature. This packet frame token is used instead of the packet frame length in downstream tasks until the last reconstruction step when the packet frame length is predicted, so generated traffic mimics network traffic from user activities occurring on IoT devices. Since packet-level signatures are specific to IoT devices, this specializes IoTFlowGenerator to synthesize IoT specific traffic and differentiates IoTFlowGenerator from generic network packet generators.


\subsubsection{Packet Timings across Multiple Magnitudes} \label{dimruddurations}
In our dataset, the duration between any two consecutive packets in a flow may vary wildly for any given device type. For some instances, the time between two consecutive packets may may be less than a thousandth of a second, while a period of inactivity may see the time between packets of minutes or more. Intuitively, if we built a model to predict packet durations directly, our auto encoder loss function will completely ignore accurately predicting smaller values, while a log transformed input will skew predicting the larger values. We bin packet durations into partitions using clustering, so varying magnitudes are preserved. Our downstream tasks predict a duration partition instead of directly predicting the duration as a continuous value.

\subsection{Finding Discrete Packet Representations}
The motivation behind finding discrete packet header representations is to reduce the dimensionality of the feature space so that the core sequence generation algorithm can learn and produce sequences at higher level. We first compute a feature vector for each packet in a traffic flow numerically encoding each of the packet's features. We train a vector quantized sequence transformer auto encoder or VQ-STAE to perform both sequence discretization and reconstruction for traffic windows per device. We then use the encoder from the trained VQ-STAE to transform packet feature vectors into discrete representations per packet - the resulting discrete sequences are learned via SeqGAN.


\subsection{Generation} \label{genpackettokens}
In IoTFlowGenerator, SeqGAN is responsible for generating traffic windows in the discrete space. SeqGAN is able to generate discrete sequences out of the box. Sometimes, the generator encounters mode collapse, a failure case for GANs where the generator learns to produce samples with extremely low variety. We counteract this issue by checking the generated samples for low variety before using them in downstream tasks.

\subsection{Reconstruction} \label{gendur}
IoTFlowGenerator uses the decoder in the previously trained VQ-STAE to map synthetic discrete sequences generated from SeqGAN back to the feature space for each device (See Figure~\ref{fig:VQSTAE}).  

\par
After we generate duration tokens in the feature space, our method for mapping duration tokens to actual duration values is simple. Each token represents a cluster of duration values partitioned from the training data. When we need to map a token to a real duration value, we randomly select a duration value from the cluster corresponding to the token, and introduce a small amount of random noise so we don't have exactly matching duration values which is extremely rare in the real world. Consider a noise element - a positive real number $e$, and original duration $q$, the final duration $q'$ will be a random number between $q - e$ and $q + e$. $e$ is calculated as a fixed fraction of $q$, and we use $e = q/10$. The target element in the traffic window is then augmented with the duration value.

\par
The feature space does not contain explicit packet frame lengths and instead contains a unique integer encoding we call a packet frame token representing a position in a signature. We map this integer encoding back into a packet frame length for each packet using a multi-layer perception deep learning model with a sliding window on the sequence, and a look behind on previous predictions as well. The inputs for the model are the packet frame token sequence as well as the discrete packet representation. An input for random noise is also integrated for each model to produce a variety of outputs even when the inputs to the model are identical. The deep learning model trains on the real packet frame lengths, the corresponding packet frame tokens and the discrete sequence generated by the encoder in the VQ-STAE. During reconstruction, the deep learning model predicts the synthetic packet frame length using the packet frame token generated by the decoder of the VQ-STAE and the synthetic discrete sequences generated from SeqGAN. The model has hyperparameters that define the size of the sliding window and the look behind window. IoTFlowGenerator trains the models using categorical cross entropy loss function and the ADAM optimizer.

\subsubsection{Packet Fuzzing}
Once the packet metadata flows are reconstructed, we need to synthesize network packets with encrypted payloads where we utilize Scapy. IoTFlowGenerator relies on the scapy's fuzz function, which is able to change any default value that is not to be calculated (like checksums) by an object whose value is random and whose type is adapted to the field, to produce a random payload field, which represents as an encrypted payload, and other packet fields not defined by our synthetic metadata.

\section{Experimental Results}
\label{sec:results}


\subsubsection{IoT Dataset}
This dataset~\cite{aaltodata} which was published by Aalto University represents the traffic emitted during the setup of 31 smart home IoT devices of 27 different types (4 types are represented by 2 devices each). For each tested device, the typical device setup process was repeated n = 20 times, and after each testing round, a hard reset to default factory settings of the tested device was performed. The data is in the form of PCAP files; one PCAP file exists for each setup process and captured traffic flow. We select 18 of these devices which have at least 20 TCP payload containing packets per packet capture for displaying results in this paper. The other device types did not contain enough data for interesting analysis. We analyze the list of IoT devices used and their supporting connectivity technologies in Appendix Table \ref{tableConnect}.

\subsection{Evaluating Generator Performance}
In this section, we present methods to evaluate the fidelity of our synthetically generated traffic by testing how well our generated data can fool the adversary AI models. Previously, authors have defined tasks to evaluate synthetically generated data against adversary models \cite{deeplearninghoneydata}. We use one of these tasks with our defined adversary type for synthetic traffic evaluation.

\subsubsection{Baseline Comparison}
We evaluate the synthetic traffic flows generated by DoppelGANger as a baseline algorithm for IoTFlowGenerator baselining. Our generation process for DoppelGANger is as follows. We utilize Wireshark to extract packet features and assign a direction for each packet - using some of the same data pre-processing steps we use for IoTFlowGenerator. We extract features to input into DoppelGANger. The first feature is the packet length, which is a categorical variable - a separate category is used for each unique packet length. The second feature is the direction, which is a categorical variable with two categories, incoming or outgoing. We also include as features the one-hot encoded protocol configuration and source and destination port. The final feature is the rescaled (min max normalized) duration, or the time interval until the next packet arrives.
We run the generation process once per device type and  evaluate the traffic as a baseline for our own generation process IoTFlowGenerator.

\begin{table}[ht]
\caption{Synthetic Data-Aware Adversary Classification Accuracy}
\begin{tabular}{ |p{3.3cm}|l|l| } 
 \hline
 \textbf{Device} & \textbf{IoTG} & \textbf{DG} \\
    \hline
    D-LinkCam & \textbf{72.4\%} & 100.0\% \\
    \hline
    D-LinkDoorSensor &  \textbf{59.9\%} & 99.3\% \\
    \hline
    D-LinkHomeHub & \textbf{64.8\%} & 95.2\% \\
    \hline
    D-LinkSensor & \textbf{64.6\%} & 97.2\% \\
    \hline
    D-LinkSiren & \textbf{71.2\%} & 100.0\% \\
    \hline
    D-LinkSwitch & \textbf{63.2\%} & 95.4\% \\
    \hline
    D-LinkWaterSensor & \textbf{62.4\%} & 93.1\% \\
    \hline
    EdimaxPlug1101W & \textbf{59.8\%} & 91.7\% \\
    \hline
    EdnetGateway & \textbf{61.3\%} & 90.5\% \\
    \hline
    HomeMaticPlug & \textbf{61.5\%} & 85.3\% \\
    \hline
    HueBridge & \textbf{61.4\%} & 96.4\% \\
    \hline
    HueSwitch & \textbf{72.6\%} & 94.1\% \\
    \hline
    MAXGateway & \textbf{62.6\%} & 91.0\% \\
    \hline
    WeMoInsightSwitch & \textbf{76.8\%} & 92.9\% \\
    \hline
    WeMoInsightSwitch2 & \textbf{57.5\%} & 93.9\% \\
    \hline
    WeMoLink & \textbf{59.3\%} & 99.3\% \\
    \hline
    WeMoSwitch & \textbf{58.9\%} & 100.0\% \\
    \hline
    Withings & \textbf{60.1\%} & 82.0\% \\
    \hline
\end{tabular}
\label{table2}
\end{table}


\subsection{Cyber Deception for a Synthetic Data-Aware Adversary}
In this scenario, the adversary has access to both labeled synthetic data and real traffic flows during training, which may help the adversary in distinguishing synthetic data/flows. The knowledge of synthetic samples may occur in scenarios where the adversary was fooled into accepting some synthetic data previously, determined the data was synthetic and he/she may start analyzing other stored data files to authenticate them. \textit{The attacker employs classification models to predict whether the traffic flow is real or fake}. The classifiers are trained to learn 2 labels, one class for a real classification, and one class for synthetic classification on a training set. The adversarial ML models are trained to discriminate between real/fake traffic of a target device.
We computed the cross validation accuracy score for each device type against each adversarial ML model trained and validated on an equal amount of real and synthetic traffic. We did this for each device type. We randomly elected an equal number of fixed length traffic windows per device type for evaluation from the real data set. For the absolute best performance in synthetic data generation, \textit{the adversary at most correctly classifies 50 percent of the data points} (i.e., random guess). \textit{If the accuracy is close 50\%, it suggests that the synthetic data is indistinguishable from the real data}.
Table \ref{table2} shows cross validation accuracy results for select devices in the Aalto dataset (IoT setup traffic). These results indicate that IoTFlowGenerator performs significantly better than DoppelGANger across both evaluation tasks.

\section{Conclusion}
\label{sec:conclusion}
We presented a generation technique which can generate synthetic IoT traffic flows, such that strong adversary evaluation models will mistakenly identify them as real network traffic flows for the intended IoT device. We find that, in terms of the adversary model we used, the real and synthetic traffic flows from IoTFlowGenerator are relatively indistinguishable and our IoTFlowGenerator generation technique beats an existing state of art GAN-based network data generation technique in our adversary model evaluation task for all devices across the dataset we use.

\section{ Acknowledgments}
The research reported herein was supported in part by NSF awards  OAC-1828467, OAC-2115094 and ARO award W911NF-17-1-0356.
\bibliographystyle{flairs}
\bibliography{samples}

\newpage

\appendix
\section{Appendix}

\subsection{Algorithms}
\begin{algorithm}
\SetAlgoLined
\KwIn{$F$ - packet length and direction flows, $d$ - distance threshold, $s$ - minimum samples threshold, $min$ - minimum signature size, $max$ - maximum signature size}
\KwOut{$P$ extracted signatures}
 \For{$i = min;\ i <= max;\ i = i + 1$}
 {
 $A$ $\leftarrow$ []\;
 \ForEach{flow $f \in F$}
 {
 $N$ $\leftarrow$ all subarrays of $F$ with size $i$\; \label{sub}
 \ForEach{$n \in N$} {
 $A$.insert($n$)\;
 }
 }
 $C$ $\leftarrow$ DBScan($A$, $d$, $s$)\;
 $S$ $\leftarrow$ []\;
 \ForEach{$D \in C$} {
 $T$ $\leftarrow$ []\;
 \For{$j = 0;\ j < i;\ j = j + 1$} {
  $U$ $\leftarrow$ []\;
  \ForEach{$d \in D$} {
  $U$.insert($D$)\;
  }
  $T$.insert((Minimum($U$), Maximum($U$))\; \label{patternrange}
 }
 $S$.insert($T$)\;
 }
 $P$[$i$] $\leftarrow$ $S$\;
 }
 \Return $P$\;
 \caption{Signature Extraction}\label{patternextraction}
\end{algorithm}

\begin{table*}[htb]
\caption{List of IoT devices used and their supported connectivity technologies}
\resizebox{\textwidth}{!}{
\begin{tabular}{ l l c c c c c}
  {\bf Identifier} & {\bf Device Model} & {\bf WiFi} & {\bf ZigBee} & {\bf Ethernet} & {\bf Z-Wave} & {\bf Other} \\
  \hline
  D-LinkCam & D-Link HD IP Camera DCH-935L & \checkmark &  &  &  & \\
    D-LinkDoorSensor & D-Link Door \& Window sensor & & & & \checkmark & \\
    D-LinkHomeHub &  D-Link Connected Home Hub DCH-G020 & \checkmark & & \checkmark & \checkmark & \\
    D-LinkSensor &  D-Link WiFi Motion sensor DCH-S150 & \checkmark &  & & & \\
    D-LinkSiren &  D-Link Siren DCH-S220 & \checkmark &  & & & \\
    D-LinkSwitch & D-Link Smart plug DSP-W215 & \checkmark &  & & & \\
    D-LinkWaterSensor & D-Link Water sensor DCH-S160 & \checkmark &  & & & \\
    EdimaxPlug1101W &  Edimax SP-1101W Smart Plug Switch & \checkmark &  & & & \\
    EdnetGateway &  Ednet Living Starter Kit Power Gateway & \checkmark &  & & & \checkmark \\
    HomeMaticPlug & Homematic pluggable switch HMIP-PS  & &  & & & \checkmark \\
    HueBridge & Philips Hue Bridge model 3241312018  & & \checkmark  & \checkmark & & \\
    HueSwitch &  Philips Hue Light Switch PTM 215Z  & & \checkmark  & \checkmark & & \\
    MAXGateway & MAX! Cube LAN Gateway for MAX! Home automation sensors & & & \checkmark & & \checkmark \\
    WeMoInsightSwitch & WeMo Insight Switch model F7C029de & \checkmark &  & & & \\
    WeMoInsightSwitch2 & WeMo Insight Switch model F7C029de & \checkmark &  & & & \\
    WeMoLink & WeMo Link Lighting Bridge model F7C031vf & \checkmark & \checkmark  & & & \\
    WeMoSwitch & WeMo Switch model F7C027de & \checkmark &  & & & \\
    Withings & Withings Wireless Scale WS-30 & \checkmark &  & & & \\
\end{tabular}
}
\label{tableConnect}
\end{table*}

\end{document}